\documentclass[twocolumn,showpacs,preprintnumbers,amsmath,amssymb]{revtex4}

\usepackage{graphicx}
\usepackage{dcolumn}
\usepackage{bm}

\begin{document}


\title{
NMR investigation of spin correlations in BaCo$_{2}$As$_{2}$}

\author{K.\ Ahilan}
\affiliation{Department of Physics and Astronomy, McMaster University, Hamilton, Ontario L8S4M1, Canada} 
\author{T.\ Imai}
\affiliation{Department of Physics and Astronomy, McMaster University, Hamilton, Ontario L8S4M1, Canada} 
\affiliation{Canadian Institute for Advanced Research, Toronto, Ontario M5G1Z8, Canada} 
\author{A.\ S.\ Sefat}
\affiliation{Materials Science and Technology Division, Oak Ridge National Laboratory, TN 37831, USA}
\author{F.\ L.\ Ning}\email{e-mail: ningfl@zju.edu.cn}%
\affiliation{Department of Physics, Zhejiang University, Hangzhou 310027, P. R. China}

\begin{abstract}
We use NMR techniques to investigate the magnetic properties of BaCo$_{2}$As$_{2}$ single crystals, the non-superconducting end member of the Co-substituted  iron-pnictide high $T_c$ superconductor Ba(Fe$_{1-x}$Co$_{x}$)$_{2}$As$_{2}$ with $x=1$.  We present $^{75}$As NMR evidence for enhancement of low frequency spin fluctuations below $\sim100$\ K.  This enhancement is accompanied by that of static uniform spin susceptibility at the wave vector ${\bf q}$ = ${\bf 0}$, suggesting that the primary channel of the spin correlations is ferromagnetic rather than antiferromagnetic.  Comparison between the NMR Knight shift $^{75}K$ and bulk susceptibility $\chi_{bulk}$ data uncovers the presence of two separate components of spin susceptibility with distinct temperature dependences, presumably because multiple electronic bands crossing the Fermi energy play different roles in the electronic properties of BaCo$_{2}$As$_{2}$
\end{abstract}

\maketitle

\section{\label{sec:level1}Introduction}
An intriguing aspect of the iron-based high $T_c$ superconductor LaFeAsO$_{1-x}$F$_{x}$ ($T_{c}\sim28$\ K) \cite{Kamihara} and related compounds such as Ba(Fe$_{1-x}$Co$_{x}$)$_{2}$As$_{2}$ ($T_{c}\sim 22 $\ K) \cite{Sefat1} is their flexible nature in substituting atoms.  For example, one can systematically alter the electronic properties of Ba(Fe$_{1-x}$Co$_{x}$)$_{2}$As$_{2}$ by increasing the Co content from the Spin Density Wave (SDW) ordered parent phase BaFe$_{2}$As$_{2}$ ($T_{SDW}=135$\ K) \cite{Jorendt3}, the superconducting phase Ba(Fe$_{0.92}$Co$_{0.08}$)$_{2}$As$_{2}$ ($T_{c}\sim 22$\ K) \cite{Sefat1}, and to the overdoped non-superconducting phase with $x\sim 0.15$ or greater \cite{Ni, Chu}.  When the Co concentration exceeds $x\sim 0.15$, the Fermi surface nesting effects between the hole bands at the center of the Brillouin zone and the electron bands at the edge of the Brillouin zone diminish \cite{Sekiba}.  In addition, the enhancement of antiferromagnetic spin fluctuations (AFSF) observed for the superconducting phase \cite{Ning1, Ning2} is suppressed in the overdoped region above $x\sim 0.15$ \cite{Ning3}, suggesting a link between the nesting effects, AFSF, and the superconducting mechanism.  Recalling that KFe$_{2}$As$_{2}$ with one less electron within the FeAs layers is a low $T_c$ superconductor \cite{Jorendt3}, a natural question to ask is, what happens if one keeps increasing the Co content $x$, and reaches BaCo$_{2}$As$_{2}$?  Would the extra 3d electron at Co sites completely alter the electronic properties of  BaCo$_{2}$As$_{2}$ from that of BaFe$_{2}$As$_{2}$?

We attempted to address these questions by synthesizing BaCo$_{2}$As$_{2}$ single crystals, and measuring their bulk properties \cite{SefatCo}.  The in-plane electrical resistivity $\rho_{ab}$ shows a metallic behavior with a quadratic $T^{2}$ temperature dependence below $\sim 60$\ K, with a modestly large resistivity coefficient $A_{ab} \sim 2 \times 10^{-3}$ $\mu \Omega$ cm$^{2}$.  The bulk susceptibility $\chi_{bulk}$ undergoes a strong enhancement toward $T=0$.   In the $T=0$ limit, the bare magnetic susceptibility data measured along the c-axis, $\chi_{bulk}\sim 5.4 \times 10^{-3}$ emu/mol, is enhanced by a factor of $\sim 23$ over the results of LDA  (local density approximation) calculations, $\chi_{LDA}\sim 2.4 \times 10^{-4}$ emu/mol \cite{SefatCo}. (We define the magnetic susceptibility for the formula unit throughout this paper.)  The linear specific heat coefficient $\gamma \sim 20.5$\ (mJ/K$^{2}$ mol-Co) also implies a modest mass renormalization of $\sim 2$ \cite{SefatCo}.  LDA results satisfy the mean-field Stoner criterion of ferromagnetism, and found a stable ferromagnetic ground state with a sizable ordered moment, 0.42$\mu_{B}$/Co, even though BaCo$_{2}$As$_{2}$ exhibits no evidence for a ferromagnetic long range order in bulk physical properties \cite{SefatCo} or microscopic NMR data, as shown below.  The LDA results therefore hint the presence of strong ferromagnetic spin correlations \cite{SefatCo}.  Recent ARPES (Angle Resolved Photo Emission Spectroscopy) measurements indeed showed that an electron pocket rather than a large hole pocket constitutes the primary Fermi surface centered around the $\Gamma$ point, eliminating the possibility of the quasi-nesting effects of the Fermi surface that tend to favor antiferromagnetic correlations \cite{Xu, Dhaka}.  On the other hand, by merely replacing Ba$^{2+}$ ions in BaCo$_{2}$As$_{2}$ with Sr$^{2+}$ ions, it is demonstrated that SrCo$_{2}$As$_{2}$ has stripe antiferromagnetic correlations, and $\chi_{bulk}$ decreases toward $T=0$ \cite{Pandey, Jayasekara}.  The contrasting behavior between BaCo$_{2}$As$_{2}$ and SrCo$_{2}$As$_{2}$ despite their analogous structures suggests that intricate details of the crystal structure, bond angles, crystal fields, and probably the electronic band structure affect their physical properties. 

In this article, we will attempt to shed new light on the physical properties of BaCo$_{2}$As$_{2}$ through a microscopic NMR investigation.  From the measurements of $^{75}$As NMR Knight shift $^{75}K$ and nuclear spin-lattice relaxation rate $1/T_{1}$, we probe the nature of electronic correlations.  Our results suggest that ferromagnetic spin fluctuations indeed begin to grow below about 100\ K.  On the other hand, comparison between $^{75}K$ and $\chi_{bulk}$ data reveals that there is a large temperature independent component in $\chi_{bulk}$ which has only a small contribution to $^{75}K$.  We therefore conclude that multiple 3d bands play different roles in BaCo$_{2}$As$_{2}$.

The rest of this paper is organized as follows.  In section II, we will briefly describe the experimental procedures.  Results and discussions, and comparison with earlier NMR works on various compositions of Ba(Fe$_{1-x}$Co$_{x}$)$_{2}$As$_{2}$ will be detailed in section III.  We will summarize and conclude in Section IV.

\section{\label{sec:level1}Experimental}
We grew BaCo$_{2}$As$_{2}$ single crystals out of CoAs flux.  The details of the growth procedures and bulk characterization are described in \cite{SefatCo}.  We carried out NMR measurements by assembling several pieces of small single crystals on a sample holder, and applying an external magnetic field along the aligned crystal c-axis.  Typical dimensions of the crystals were $\sim1$ mm $\times$ 1mm $\times$ 0.1 mm.  We estimate the total mass of our NMR sample as a few mg.  We employed a standard pulsed NMR spectrometer for spin echo measurements.  

\section{\label{sec:level1}Results and Discussions}
\subsubsection{$^{75}$As NMR lineshapes and $^{75}\nu_{Q}^{c}$}
 In Fig.\ 1, we present representative $^{75}$As NMR lineshapes observed at various temperatures in a magnetic field of $B=7.734$~T applied along the aligned crystal c-axis.   $^{75}$As nuclear spin is $I=3/2$, hence we observe three distinct transitions at 290\ K; the large sharp peak near 56.69~MHz arises from the $I_{z}=+1/2$ to $-1/2$ central transition, while two additional broad peaks located near 56.14 and 57.22~MHz are from the $I_{z}=\pm3/2$ to $\pm1/2$ satellite transitions.  The growing width of the central peak at lower temperatures is primarily caused by a distribution of the Knight shift $^{75}K$.

 \begin{figure}
\includegraphics[width=3in]{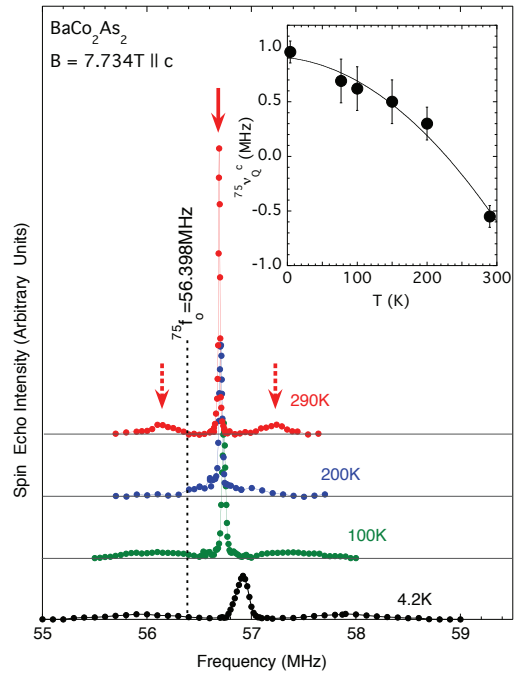}\\
\caption{\label{Fig.1} (Color Online) Representative $^{75}$As NMR lineshapes observed in a magnetic field $B = 7.734$\ T applied along the aligned c-axis of single crystals.  A downward arrow marks the central transition at 290\ K, while dashed arrows mark the satellite transitions split from the central transition by $|^{75}\nu_{Q}^{c}| \sim 0.55$\ MHz.  For clarity, we shifted the origin of the horizontal axis for different temperatures.  The vertical dotted line represents the bare Zeeman frequency $^{75}f_{o}=56.398$\ MHz, where we expect to observe the central peak if the Knight shift $^{75}K$ was vanishingly small.  Inset: The temperature dependence of the c-axis component of the $^{75}$As nuclear quadrupole frequency $^{75}\nu_{Q}^{c}$.  For clarity, we assume  $^{75}\nu_{Q}^{c} > 0$ at 4.2~K.  The solid curve is a guide for the eyes.  }  
\end{figure}
 
The splitting $^{75} \nu_{Q}^{c}$ between the central and two satellite transitions is caused by the nuclear quadrupole interaction.  The broader lineshape of the satellites reflects a sizable distribution of the c-axis component of the nuclear quadrupole frequency $^{75} \nu_{Q}^{c}$, possibly due to the presence of defects or vacancy sites.  $^{75}\nu_{Q}$ is a traceless tensor, i.e. $^{75}\nu_{Q}^{a}+ ^{75}\nu_{Q}^{b} + ^{75}\nu_{Q}^{c} =0$, and is proportional to the EFG (Electric Field Gradient) tensor at the observed $^{75}$As site.  NMR measurements alone do not allow us to determine the sign of EFG and $^{75} \nu_{Q}^{c}$.  The observed magnitude of $|^{75} \nu_{Q}^{c} | = 0.96$\ MHz at 4.2\ K is comparable to $|^{75} \nu_{Q}^{c} | = 2.4$\ MHz observed for the optimally superconducting Ba(Fe$_{0.9}$Co$_{0.1}$)$_{2}$As$_{2}$ ($T_{c}=22$\ K) \cite{Ning1}.  

We summarize the temperature dependence of $^{75} \nu_{Q}^{c}$ in the inset to Fig.\ 1, assuming that  $^{75} \nu_{Q}^{c}$ has a positive sign at 4.2~K.  We emphasize that this assumption is made genuinely for convenience, to avoid confusions in the following discussions.  $|^{75} \nu_{Q}^{c}|$ gradually decreases with temperature up to $\sim 200$~K, where the satellite transitions are superposed with the central transition.  This smooth temperature dependence is presumably caused by the thermal expansion of the lattice, which tends to decrease the magnitude of the lattice contribution \cite{Cohen} to $^{75} \nu_{Q}^{c}$.    At $\sim 290$\ K,  $|^{75} \nu_{Q}^{c}|=0.5$\ MHz becomes sizable, and we can observe distinct satellite peaks again. The implicit assumption made in our plot in the inset to Fig.~1 is that there is a negative, on-site ionic contribution to $^{75} \nu_{Q}^{c}$ \cite{Cohen}.  Our experimental results do not rule out an alternative possibility that the sign of $^{75} \nu_{Q}^{c}$ remains the same, and  $|^{75} \nu_{Q}^{c}|$ increases from $\sim 200$~K to $\sim 290$\ K.  It is also interesting to recall that $|^{75} \nu_{Q}^{c}|$ in the iso-structural SrCo$_{2}$As$_{2}$ is as large as $\sim 10$~MHz \cite{Pandey}.  An order of magnitude difference in $|^{75} \nu_{Q}^{c}|$ suggests that the local charge and/or lattice environment at the As sites is markedly different between the two materials, which may be related to the fact that BaCo$_2$As$_2$ is ferromagnetically correlated, as shown below, while SrCo$_2$As$_2$ is antiferromagnetically correlated \cite{Jayasekara}.
 
 \subsubsection{NMR Knight shift $^{75}K$}
 In Fig.~2(a), we present the temperature dependence of $^{75}$As NMR Knight shift, $^{75}K = (^{75}f - ^{75}f_{o})/^{75}f_{o}$, where $^{75}f$ is the observed peak frequency of the central transition, $^{75}f_{o}$ is the bare Zeeman frequency $^{75}f_{o}=^{75}\gamma_{n} B = 56.398$~MHz, and the $^{75}$As nuclear gyromagnetic ratio is $^{75}\gamma_{n}/2\pi=7.2919$~MHz/T.  The positive frequency shift from $^{75}f_{o}$ to $^{75}f$ is caused by the Knight shift $^{75}K$ through hyperfine interactions with electron spins polarized by the applied field $B$.  For comparison, we also present $^{75}K$ reported earlier for Ba(Fe$_{1-x}$Co$_{x}$)$_{2}$As$_{2}$ for representative compositions with $x=0$ (SDW ordered below $T_{SDW}\sim 135$~K), $x=0.08$ (superconducting below $T_{c}\sim 25$~K), and $x=0.26$ (overdoped paramagnetic metal) \cite{Ning3}.  
 
 Quite generally, $^{75}K$ may be represented as 
\begin{equation}
          ^{75}K = \Sigma_{j} \left( \frac{A_{HF}^{(j)}}{N_{A}\mu_{B}} \right) \cdot \chi_{spin}^{(j)} + ^{75}K_{chem},
\label{2}
\end{equation}
where $N_{A}$ is Avogadro's number, and $^{75}K_{chem}$ is the chemical shift.  The latter may be considered temperature independent unless we are dealing with unusual circumstances, such as a change of the valence at the Co sites or a structural phase transition.  $\chi_{spin}^{(j)}$ is the j-th component of the spin susceptibility, and $A_{HF}^{(j)}$ is the hyperfine coupling between the observed nuclear spin and the electrons contributing to $\chi_{spin}^{(j)}$.  The superscript $j$ represents the band or orbital indices, accounting for different contributions with potentially distinct temperature dependences.   We recall that all of the five 3d Fe and Co orbitals may contribute to multiple electronic bands in Ba(Fe$_{1-x}$Co$_{x}$)$_{2}$As$_{2}$ and related compounds \cite{Singh, Kuroki, Graser, SefatCo, Xu, Pandey}; in principle $\chi_{spin}^{(j)}$ associated with different bands may exhibit different temperature dependences.  Analogous situation was encountered for Sr$_2$RuO$_4$ \cite{ImaiRu}.  In what follows, we need to keep in our mind that $^{75}K$ is an average of various contributions of $\chi_{spin}^{(j)}$ weighted by $A_{HF}^{(j)}$. 

\begin{figure}[b]
\includegraphics[width=3.5in]{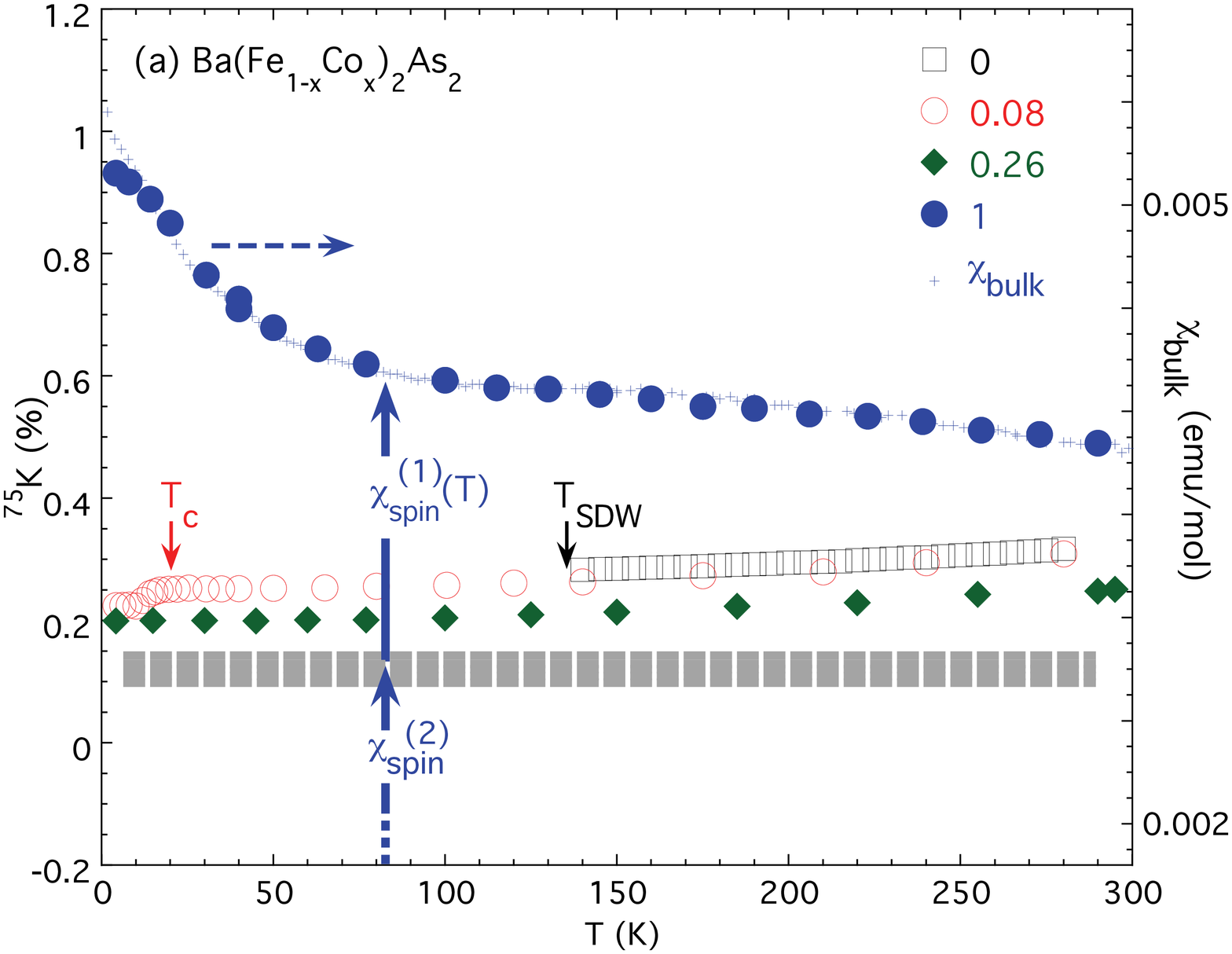}\\
\includegraphics[width=3.5in]{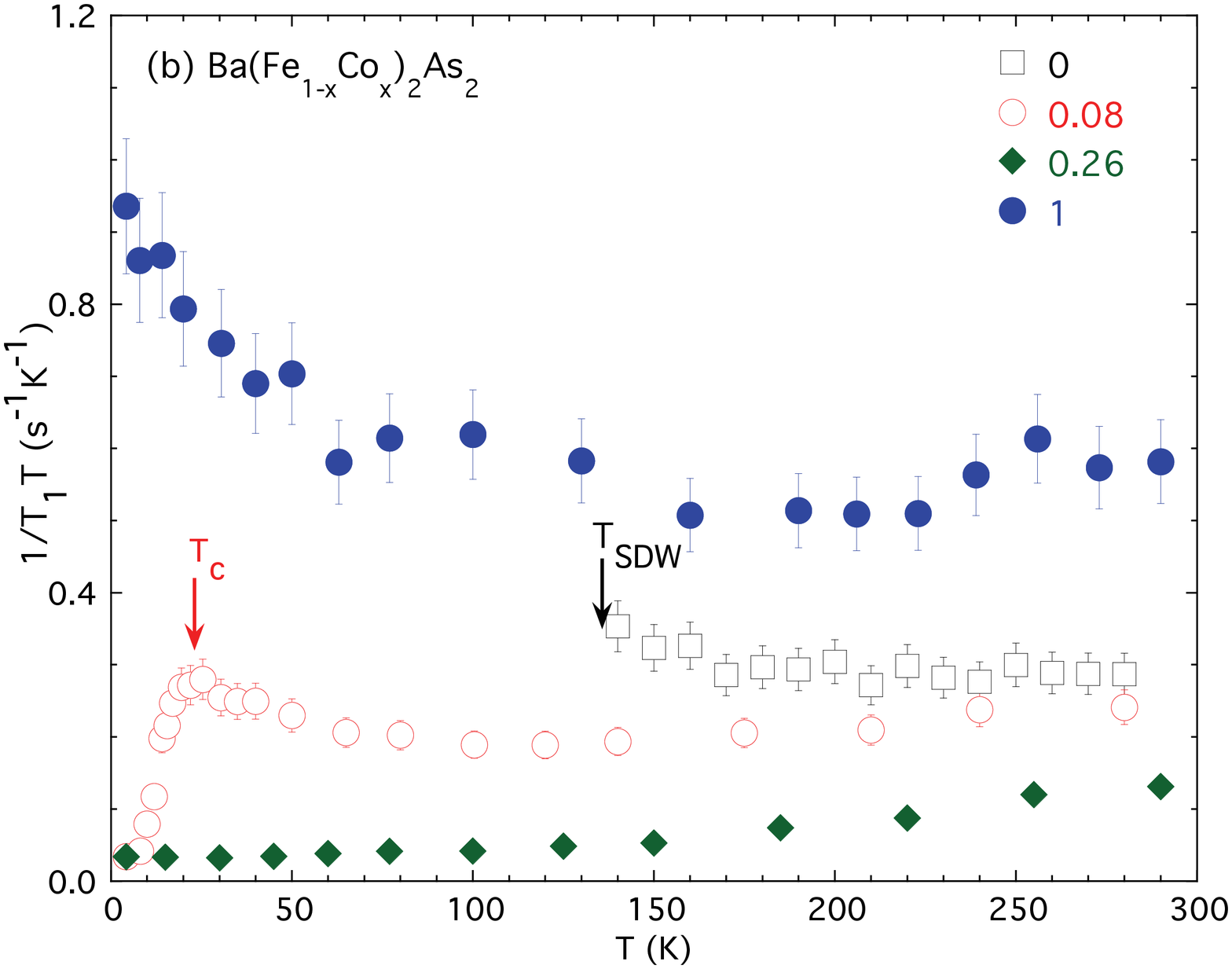}\\
\caption{\label{Fig.2} (Color Online) (a) (Left axis) Comparison of the $^{75}$As NMR Knight shift $^{75}K$ observed for BaCo$_{2}$As$_{2}$ ($\bullet$) and Ba(Fe$_{1-x}$Co$_{x}$)$_{2}$As$_{2}$ for representative compositions: ($\square$) the SDW phase $x=0$ with $T_{SDW}=135$~K, ($\circ$) the superconducting phase $x=0.08$ with $T_{c}=22$~K, and ($\blacklozenge$) the over-doped non-superconducting metallic phase with $x=0.26$.  Vertical arrows mark the superconducting transition temperature $T_{c}$ for $x=0.08$ and SDW transition temperature $T_{SDW}$ for $x=0$.  (Right axis) The bulk susceptibility $\chi_{bulk}$ (measured in 0.1 T) of BaCo$_{2}$As$_{2}$ matched with $^{75}K$.  The wide horizontal dashed line represents the approximate value of $^{75}K_{chem} = 0 \sim 0.2$\%, which corresponds to the temperature independent contribution $\chi_{spin}^{(2)} = 2.5 \sim 3 \times 10^{-3}$ emu/mol.  (b) Temperature dependence of $1/T_{1}T$.  The symbols are the same as in (a).
}  
\end{figure}

\begin{figure}
\includegraphics[width=3in]{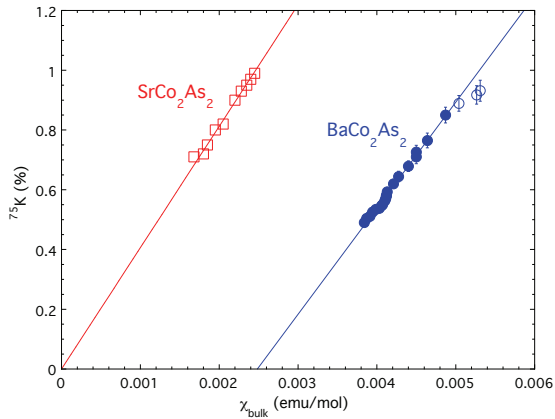}\\
\caption{\label{Fig.3}(Color Online) The Knight shift $^{75}K$ in BaCo$_{2}$As$_{2}$ plotted as a function of $\chi_{bulk}$ with temperature as the implicit parameter.  The linear fit through the data point shows the linear relation between $^{75}K$ and $\chi_{spin}^{(1)}(T)$.  Slight deviation from the linear fit at low temperatures (shown with open circles) is caused by a small impurity contribution in $\chi_{bulk}$ below $\sim 10$~K, as evident in Fig.~2(a).  For comparison, we also show the corresponding result reported for SrCo$_{2}$As$_{2}$ \cite{Pandey}.
}  
\end{figure}

We find two striking features in our $^{75}K$ data in Fig.\ 2(a).  First, the magnitude of $^{75}K$ in BaCo$_2$As$_2$ is much greater than that observed for Ba(Fe$_{1-x}$Co$_{x}$)$_{2}$As$_{2}$ with $x \leq 0.26$.  This observation is consistent with the fact that $\chi_{bulk}$  is also nearly an order of magnitude greater in BaCo$_{2}$As$_{2}$.  Second, the observed temperature dependence of $^{75}K$ is qualitatively different from that of  Ba(Fe$_{1-x}$Co$_{x}$)$_{2}$As$_{2}$.  In the latter, $^{75}K$ decreases with temperature, reflecting the common trend in all iron-based high $T_c$ superconductors, including LaFeAs(O$_{1-x}$F$_{x}$) \cite{Ahilan, Nakai}, FeSe \cite{ImaiFeSe}, and K$_{x}$Fe$_{2-y}$Se$_{2}$ \cite{Yu, Kotegawa, Torchetti}.  Instead, $^{75}K$ in BaCo$_{2}$As$_{2}$ increases with decreasing temperature  from 300\ K, reaches a plateau below $\sim 200$\ K, followed by a rapid growth below $\sim 100$\ K.   

Also plotted in Fig.\ 2(a) using the right axis is the temperature dependence of the bulk magnetic susceptibility $\chi_{bulk}$ measured in 0.1 Tesla applied along the c-axis \cite{SefatCo}.  The observed temperature dependence is qualitatively the same as that of $^{75}K$.  The upturn observed below $\sim 10$\ K only for $\chi_{bulk}$ is presumably caused by a small impurity contribution obeying the Curie's law, $\sim 1/T$, because it is absent in the $^{75}K$ data.  We note that $^{75}K$ is generally immune from the presence of impurity phases, and saturates below $\sim 20$~K.  In Fig.~3,  we plot $^{75}K$ as a function of $\chi_{bulk}$ with temperature as the implicit parameter.  The straight line observed in a wide range of temperature implies that the temperature dependence of $^{75}K$ is indeed dominated by the temperature dependent component of  $\chi_{bulk}$, which we refer to $\chi_{spin}^{(1)}(T)$.  
        
In the case of Ba(Fe$_{1-x}$Co$_{x}$)$_{2}$As$_{2}$, the temperature independent chemical shift $^{75}K_{chem}$ is known to be at most $\sim 0.2$\% \cite{Ning1,Kitagawa}.  Recent $^{75}$As NMR results for other compounds with Co$^{2+}$ ions, SrCo$_2$As$_2$ \cite{Pandey} and LaCoAsO \cite{Yoshimura}, also suggest that $^{75}K_{chem}$ is negligibly small, and hence the extrapolation of the linear fit of the $^{75}K$ vs. $\chi_{bulk}$ goes through the origin.    We reproduce such a linear fit  observed for SrCo$_2$As$_2$ \cite{Pandey} in Fig.~3.  In contrast with the case of SrCo$_2$As$_2$, the linear fit for BaCo$_2$As$_2$ crosses the horizontal axis at a large positive value of $\sim 2.5 \times 10^{-3}$\ emu/mol.    This term is too large to be attributed to the Van Vleck contribution, $\chi_{VV}$, which is of the order of $\chi_{VV}\sim 10^{-4}$\ emu/mol for 3d bands with a typical band width of $\sim 1$~eV.  We therefore conclude that there is another temperature independent contribution $\chi_{spin}^{(2)} \sim 2.5 \times 10^{-3}$\ emu/mol in $\chi_{bulk}$, and we may separate the overall spin contribution into two, $\chi_{bulk} \sim \chi_{spin}^{(1)}(T) + \chi_{spin}^{(2)}$.  We show these two separate contributions schematically with two vertical arrows in Fig.~2(a).  Both the temperature dependent $\chi_{spin}^{(1)}(T)$ and temperature independent $\chi_{spin}^{(2)}$ are an order of magnitude greater than the bare spin susceptibility, $\chi_{LDA}\sim 2.4 \times 10^{-4}$ emu/mol, estimated by LDA calculations \cite{SefatCo}.  

We can use Eq.\ (1) to estimate the hyperfine coupling $A_{hf}^{(1)}$ with Co spins in the band that is responsible for $\chi_{spin}^{(1)}(T)$.  From the slope in Fig.\ 3,  we estimate $A_{hf}^{(1)} = 3.9$ T/$\mu_{B}$.  This value is comparable to $A_{hf} = 4.5$ T/$\mu_{B}$ reported for SrCo$_2$As$_2$ \cite{Pandey}, and approximately twice larger than $A_{hf} = 1.9$\ T/$\mu_{B}$ reported for BaFe$_{2}$As$_{2}$ \cite{Kitagawa}.  The stronger hyperfine coupling might be an indication that Co orbitals responsible for enhanced ferromagnetic correlations hybridize very strongly with As 4s and/or 4p orbitals.  On the other hand, Fig.~2(a) and Fig.~3 suggest that the $\chi_{spin}^{(2)}$ term has a very little contribution to $^{75}K$, which in turn implies that the hyperfine coupling $A_{hf}^{(2)}$ with the Co 3d orbitals that are responsible for $\chi_{spin}^{(2)}$ is much weaker than $A_{hf}^{(1)}$.  

The electronic band structure calculations and ARPES measurements for BaCo$_{2}$As$_{2}$ revealed two new distinct features compared with the case of BaFe$_{2}$As$_{2}$: a small electron pocket near the $\Gamma$ point, which may favor ferromagnetic correlations, and a flat band, which tends to results in a large density of states (and hence a large contribution to Pauli susceptibility) \cite{SefatCo, Xu, Dhaka}.  Therefore we tentatively attribute $\chi_{spin}^{(1)}(T)$ to the electron pocket near the $\Gamma$ point, while the temperature independent $\chi_{spin}^{(2)}$ primarily to the flat portion of the band.  We cannot rule out other scenarios, and call for additional theoretical analysis of the band contributions to spin susceptibility in BaCo$_{2}$As$_{2}$.

\subsubsection{Low Frequency Spin Fluctuations As Investigated By Nuclear Spin-Lattice Relaxation Rate $1/T_{1}$}
In Fig.\ 2(b), we present the temperature dependence of $1/T_{1}T$, i.e. $^{75}$As nuclear spin-lattice relaxation rate $1/T_{1}$ divided by temperature $T$.  In general, $1/T_{1}T$ is a measure of the low-frequency component of spin fluctuations at the NMR frequency $^{75}f$  integrated over various wave vector ${\bf q}$ modes in the first Brillouin zone \cite{Moriya}.  $1/T_{1}T$ for BaCo$_{2}$As$_{2}$ is nearly a factor of $\sim 2$ larger than that observed for the paramagnetic state of BaFe$_{2}$As$_{2}$ with a SDW ground state.  This may be partly because the hyperfine couplings are twice larger in the present case, as discussed in the previous section.  

It is important to realize that the temperature dependence of $1/T_{1}T$ observed for BaCo$_{2}$As$_{2}$ below $\sim 200$\ K is qualitatively similar to that of the uniform ${\bf q}={\bf 0}$ mode of the spin susceptibility, $\chi_{spin}^{(1)}(T)$.  This finding is different from the case of Ba(Fe$_{1-x}$Co$_{x}$)$_{2}$As$_{2}$, where the growth of $1/T_{1}T$ at low temperatures is caused by antiferromagnetic spin fluctuations \cite{Ning1, Ning2, Ning3}, and hence the uniform  spin susceptibility does not grow at low temperatures.  These considerations lead us to conclude that the primary channel of the electronic correlations in BaCo$_{2}$As$_{2}$ is ferromagnetic in nature.

\section{\label{sec:level1}Summary and Conclusions}
We have reported NMR measurements of local spin susceptibility and low frequency spin fluctuations of BaCo$_{2}$As$_{2}$.  We demonstrated that both the NMR Knight shift  $^{75}K$ and $1/T_{1}T$ grow toward $T=0$ due to ferromagnetic spin correlations.  Our conclusion is in agreement with earlier results based on bulk susceptibility and specific heat data, combined with LDA calculations \cite{SefatCo}.  In addition, our results in Fig.\ 2 show that the so-called Korringa ratio, $S = \frac{4 \pi k_{B}}{\hbar} (\frac{\gamma_{n}}{\gamma_{e}})^{2} T_{1}TK^{2}$ \cite{Moriya}, is as large as $\sim$10.3 at 4.2~K.  Ignoring the potential complications arising from the quasi-two dimensionality, the multi-band nature of the electronic states, and the wave-vector dependence of hyperfine couplings, such a large Korringa ratio, $S \sim 10.3$, is indeed consistent with the presence of ferromagnetic correlations.

Our findings are in a remarkable contrast with the case of the other end member, BaFe$_{2}$As$_{2}$, with a SDW ordered ground state, and modestly Co-substituted superconducting Ba(Fe$_{1-x}$Co$_{x}$)$_{2}$As$_{2}$.  In these cases, the growth of spin fluctuations reflected on $1/T_{1}T$ is not accompanied by that of $^{75}K$, because the dominant correlation effects are antiferromagnetic at least up to $x\sim0.15$ \cite{Ning3}.   Our new NMR results are also qualitatively different from those reported for SrCo$_{2}$As$_{2}$ \cite{Pandey} with a nearly antiferromagnetic ground state \cite{Jayasekara}; both NMR Knight shift and $1/T_{1}T$ decrease with temperature toward $T=0$ in SrCo$_{2}$As$_{2}$.   It remains to be seen why the electronic properties are so different between BaCo$_{2}$As$_{2}$ and SrCo$_{2}$As$_{2}$. 

From the comparison of $\chi_{bulk}$ and $^{75}K$, we also inferred that ferromagnetic enhancement does not affect all electronic bands.  Some Co 3d electrons contribute to a temperature dependent term of spin susceptibility, which grows to $\chi_{spin}^{(1)}(T)\sim 3 \times 10^{-3}$~emu/mol at 4.2~K.   On the other hand, an equally large but temperature independent contribution, $\chi_{spin}^{(2)}\sim 2.5 \times 10^{-3}$~emu/mol, arises from Co 3d orbitals with smaller hyperfine couplings with $^{75}$As nuclear spins.  The latter implies that these Co 3d orbitals have weaker hybridization with As 4s and 4p orbitals.  Preliminary $^{59}$Co NMR measurements showed that the $^{59}$Co NMR Knight shift $^{59}K$ exhibits analogous temperature dependence as $^{75}$As NMR Knight shift $^{75}K$.  The sign of $^{59}K$ in the present case, however, is negative, rather than positive observed in Ba(Fe$_{0.9}$Co$_{0.1}$)$_{2}$As$_{2}$ \cite{Ning1}.  The reversal of the sign suggests that the hyperfine coupling of Co orbitals, and hence the nature of their hybridization, is different in the present case.   We caution that our NMR data and analysis do not entirely rule out the possibility of antiferromagnetic spin correlations associated with $\chi_{spin}^{(2)}$.  In this context, it is worth recalling that another well-known multi-band system, Sr$_2$RuO$_4$, turned out to exhibit strong antiferromagnetic spin correlations for two of the three bands, contrary to the initial expectations held by many authors \cite{Braden}.  

It is natural to speculate that  $\chi_{spin}^{(1)}$ and $\chi_{spin}^{(2)}$ arise from separate parts of the Fermi surface, such as the electron pocket near the $\Gamma$ point and the flat portion of the band with a large density of states, respectively.  Regardless of their origins, our finding implies that Hund's coupling is not strong enough in BaCo$_{2}$As$_{2}$ to force all seven 3d electrons at Co$^{2+}$ sites to exhibit an identical behavior.  In this sense, the magnetism in BaCo$_{2}$As$_{2}$ seems more itinerant in character than localized, in agreement with the fact that resistivity shows a Fermi liquid behavior below $\sim 60$~K \cite{SefatCo}.  \\

\begin{acknowledgments}
The work at McMaster was supported by NSERC and CIFAR.  The work at Oak Ridge National Laboratory was supported by the Department of Energy, Basic Energy Sciences, Materials Sciences and Engineering Division.   The work at Zhejiang was supported by National Basic Research Program of China (No.2014CB921203, 2011CBA00103), NSF of China (No. 11274268).
\end{acknowledgments}


\end{document}